# DESIGN OF THE LBNF BEAMLINE TARGET STATION*

S. Tariq†, K. Ammigan, K. Anderson, S. A. Buccellato, C. F. Crowley, B. D. Hartsell, P. Hurh, J. Hylen, P. Kasper, G. E. Krafczyk, A. Lee, B. Lundberg, A. Marchionni, N. V. Mokhov, C. D. Moore, V. Papadimitriou, D. Pushka, I. Rakhno, S. D. Reitzner, V. Sidorov, A. M. Stefanik, I. S. Tropin, K. Vaziri, K. Williams, R. M. Zwaska, Fermilab, Batavia, IL 60510, USA
C. Densham, STFC/RAL, Didcot, Oxfordshire, OX11 0QX, UK

*Abstract*

The Long Baseline Neutrino Facility (LBNF) project will build a beamline located at Fermilab to create and aim an intense neutrino beam of appropriate energy range toward the DUNE detectors at the SURF facility in Lead, South Dakota. Neutrino production starts in the Target Station, which consists of a solid target, magnetic focusing horns, and the associated sub-systems and shielding infrastructure. Protons hit the target producing mesons which are then focused by the horns into a helium-filled decay pipe where they decay into muons and neutrinos. The target and horns are encased in actively cooled steel and concrete shielding in a chamber called the target chase. The reference design chase is filled with air, but nitrogen and helium are being evaluated as alternatives. A replaceable beam window separates the decay pipe from the target chase. The facility is designed for initial operation at 1.2 MW, with the ability to upgrade to 2.4 MW, and is taking advantage of the experience gained by operating Fermilab's NuMI facility. We discuss here the design status, associated challenges, and ongoing R&D and physics-driven component optimization of the Target Station.

## INTRODUCTION

The Target Station is a central component of the LBNF Beamline and it is expected to produce the highest power neutrino beam in the world. The driving physics consideration is the study of long baseline neutrino oscillations. The initial operation of the facility will be at a beam power of 1.2 MW on the production target, however some of the initial implementation will have to be done in such a manner that operation at 2.4 MW can be achieved without major retrofitting [1]. In general, components of the Target Station which cannot be replaced or easily modified after substantial irradiation at 1.2 MW operation are being designed for 2.4 MW, mainly the shielding around the target chase and the associated remote handling equipment. Approximately 40% of the total beam power is deposited in the target chase and surrounding shielding. Experience gained from operating Fermilab's NuMI target facility [2] is being extensively employed in the design of the LBNF facility.

The Target Station design has to implement a stringent radiological protection program for the environment, workers and members of the public. The relevant radiological concerns: prompt dose, residual dose, air activation, and water activation are being extensively modelled (together with NuMI benchmarking) and the results incorporated in the system design. Tritium build-up in the shielding infrastructure has to be managed and the choice of gas in the target chase has to be selected taking into consideration the impact on air emissions plus any corrosion related issues due to the ionizing radiation. The replaceable beam window separating the target chase from the helium-filled decay pipe needs to seal reliably and adequately between these two beamline volumes.

The reference design utilizes a NuMI-style two horn system and a two-interaction length carbon fin target. Further physics-driven component optimization work has shown that a three horn focusing system with a longer target provides improvements in the neutrino flux.

Radiation damage, cooling of elements, radionuclide mitigation, remote handling and storage of radioactive components are all essential considerations for the design of the LBNF Target Station. This paper gives a snapshot of the present design status of the Target Station and discusses the associated challenges, ongoing R&D, and physics-driven optimization of the target and horn system.

## STATUS OF THE TARGET STATION DESIGN

LBNF/DUNE obtained CD-1 approval in November 2015 for the reference design. A longitudinal section through the reference LBNF Target Station design is shown in Fig. 1. It closely follows the design of the NuMI focusing system where components are loaded from the top. It includes in order of placement (1) a beryllium window that seals off and separates the evacuated primary beamline from the neutrino beamline, (2) a baffle collimator assembly to protect the target and the horns from mis-steered beam, (3) a 95 cm long target, (4) two magnetic horns. The LBNF horns operate at higher current (230 kA) and lower pulse width (0.8 ms) compared to NuMI. These elements are all located inside a heavily shielded, air-filled, air/water-cooled chamber, called the target chase, that is isolated from the decay pipe at its downstream end by a replaceable, thin, metallic window. The target chase has sufficient length and cross-section to accommodate an optimized focusing system (currently being studied). In order to mitigate potential corrosion issues and release of air-born radionuclides we are studying alternative gases (nitrogen and helium) for the target chase atmosphere, and a complete conceptual design is currently being developed for the nitrogen-filled option. The remote handling facilities in the Target Hall include a shielded work-cell for remotely exchanging radioactive components and a storage rack

* Work supported by Fermi Research Alliance, LLC under Contract No. DE-AC02-07CH11359 with the United States Department of Energy.
† tariq@fnal.gov

(morgue) for staging radioactive components. Remote handling of components is accomplished with long-reach tools, remote vision systems, and a specially modified bridge crane.

There has been significant progress on the target and horn optimization effort post CD-1, including a conceptual mechanical model of a three horn design and thermo-mechanical analyses on the first two horns. Further physics-driven optimization work continues on the horn geometries. Also, various design options for the replaceable decay pipe beam window have been investigated.

On-going studies are examining the activated air emissions for the reference design (air-filled chase) and the nitrogen-filled option. A corrosion working group was formed to better understand corrosion related issues and a proposal is in place to study the corrosion rates for the life-of-facility materials, primarily the carbon steel shielding, decay pipe and associated welds.

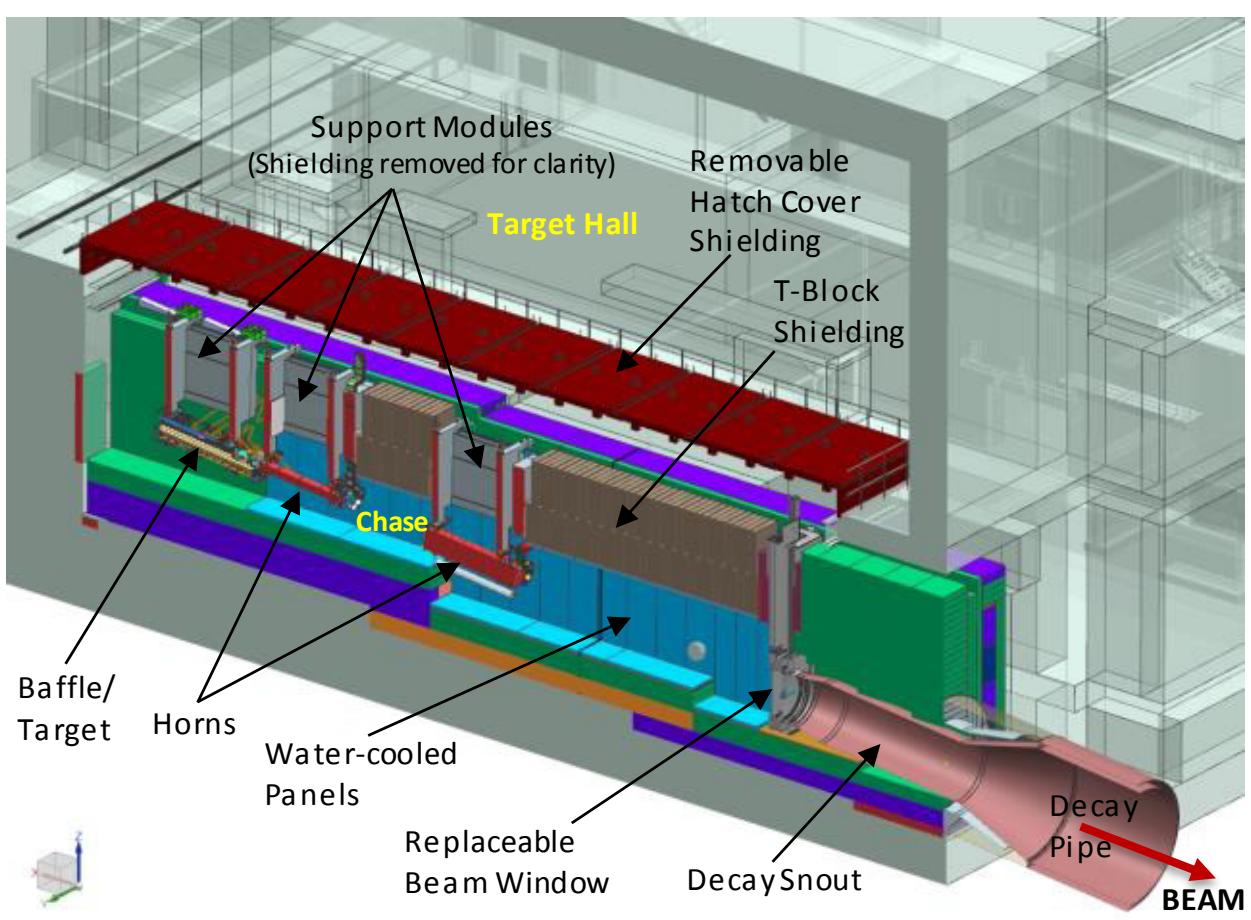

Figure 1: Target facility reference design cross-section showing, from left to right, the horn-protection baffle and target, followed by the two focusing horns, beam window, and beginning of the decay pipe. The chase is ~34 m long and ~2.1 m wide, and the height varies.

## *Target Chase Shielding & Cooling Gas Medium*

The target chase shielding design, shown in Fig. 1, is based on NuMI and designed to (1) keep the accumulated radionuclide concentration levels in the surrounding soil below standard detectable limits, (2) keep prompt radiation levels low enough for electronics in the Target Hall to have adequate lifetimes, and (3) keep residual radiation rates on top of the shield pile low enough to allow personnel access to the top of the shielding for maintenance. The shielding cannot be modified/upgraded later, and is therefore designed for 2.4 MW beam power. The shielding steel is stacked in a staggered and interlocking fashion so there are no line-of-sight cracks. Removable, specially made steel blocks called "T-blocks" are used to close off the top of the chase where the beamline components are installed, whereas custom steel blocks and slabs are used in the other areas. Special support modules are used to place and align the components inside the chase. The modules rest on carriage beams that are supported on the outside concrete walls for thermal (alignment) stability.

Half of the energy deposited in the shielding is removed by water-cooled panels placed on the inside of the chase (see Fig. 1), and the other half by a gas-cooling system that for the air-filled reference design flows 35,000 scfm (16.5 $m^3/s$) through the chase. The ultimate choice of cooling gas in the target chase will depend on the results from the on-going air release studies and corrosion work. Controlled air leaks are designed into the system and are part of the tritium mitigation strategy.

Changing to an inert gas cooling medium such as nitrogen will require a much better sealed system with minimal leak rates (~6 cfm or about 2 orders of magnitude less than that of air), due to ODH and nitrogen cost considerations. This is quite challenging to achieve while minimizing costs. The design will include the addition of a leak-tight barrier (welded stainless steel sheet metal) between the concrete walls of the target pit and the target chase shielding steel. The concrete walls (which provide the component support) will need to be maintained within a small enough temperature range in order to maintain component alignment. Also the design will require robust and repeatable leak-tight seals at all openings (such as hatch covers) plus leak tight seals at all interfaces and feedthroughs (such as stripline and utility penetrations). The supporting cross-members for the hatch covers will need to be removable (modular design) to allow different component configurations in the future. Sealing at the seams between the hatch covers and cross-beam interfaces is especially challenging.

## *Target and Magnetic Focusing Horns*

The magnetic focusing system currently being studied consists of an optimized three-horn configuration as shown in Fig. 2, and operating at 300 kA. The proton beam comes from the left and interacts with a 2 m long NuMI-style graphite finned target supported inside of horn 1. The mesons produced at the target enter the decay pipe immediately after horn 3.

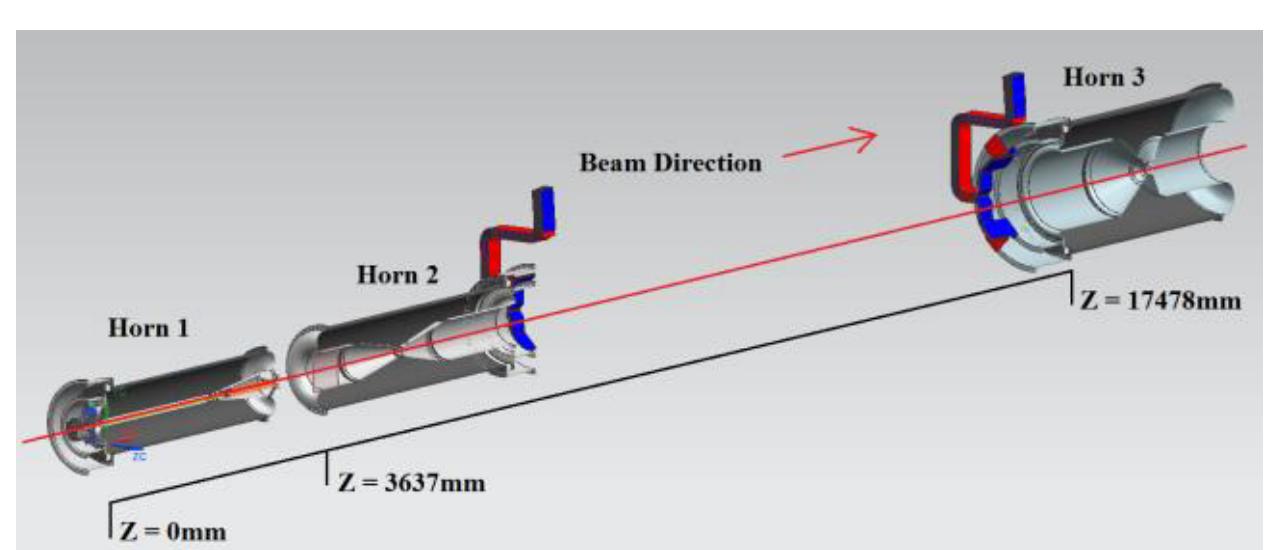

Figure 2: Three-horn configuration with power supply bus bar concepts and relative placement in beamline.

All horn conductors are constructed from 6061-T6 aluminium alloy for corrosion resistance and high strength to resist the heating effects of beam energy deposition. The integrally supported target in horn 1 is constructed from POCO Graphite grade ZXF-5Q, which is brazed to titanium water cooling lines and contained by a titanium tube.

Axial helium flow provides an inert environment while cooling the tube and downstream window.

Preliminary analysis efforts have demonstrated that the initial geometries chosen for the horn 1 and 2 inner and outer conductors are sufficient for expected temperature and stress considerations. Steady-state temperature results for horn 1 are given in Fig. 3. Optimization work continues on the three horn system, which yields a more cylindrically shaped horn 1 that tapers down on the downstream end, followed by a longer horn 2.

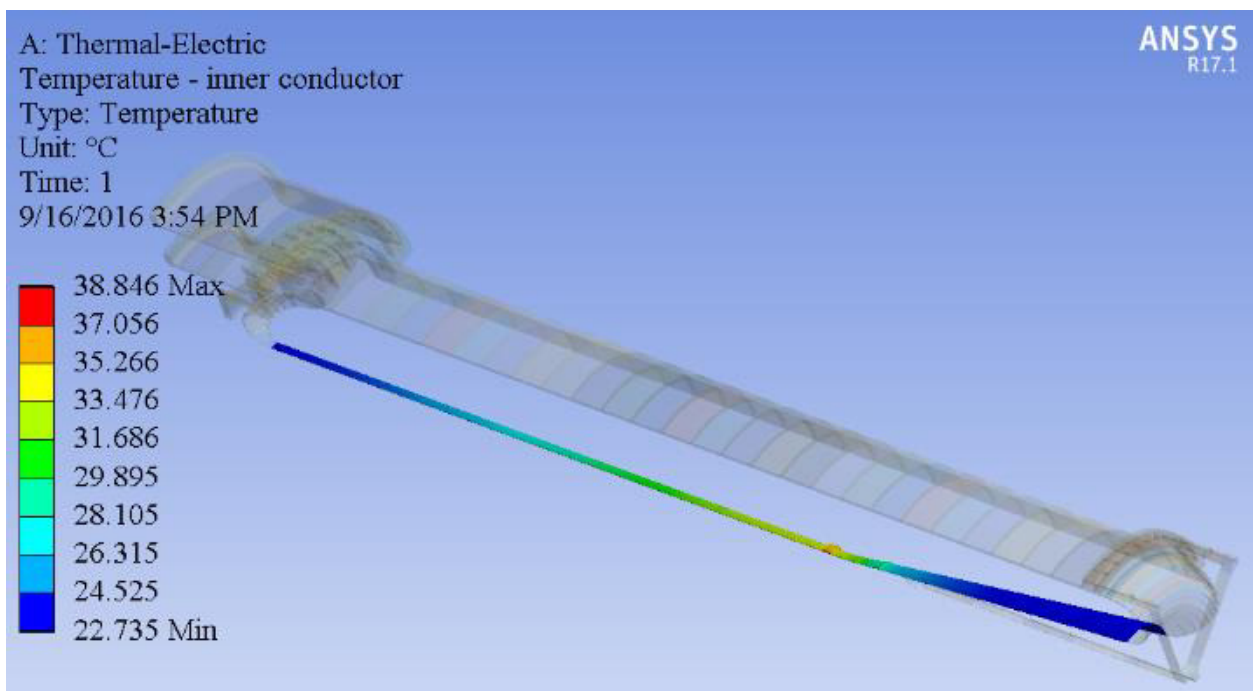

Figure 3: Preliminary steady state thermal analysis of horn 1 inner conductor.

Robust support of a 2 m long target within horn 1 and adequate cooling of its associated downstream window is challenging. Tentative design philosophy is to extend the target containment tube through the end of horn 1, and support it through the use of helium cooled titanium tubes as shown in Fig. 4. Internal diverter vanes can be used to reduce helium stagnation points at the centre of the downstream window to allow for more uniform cooling.

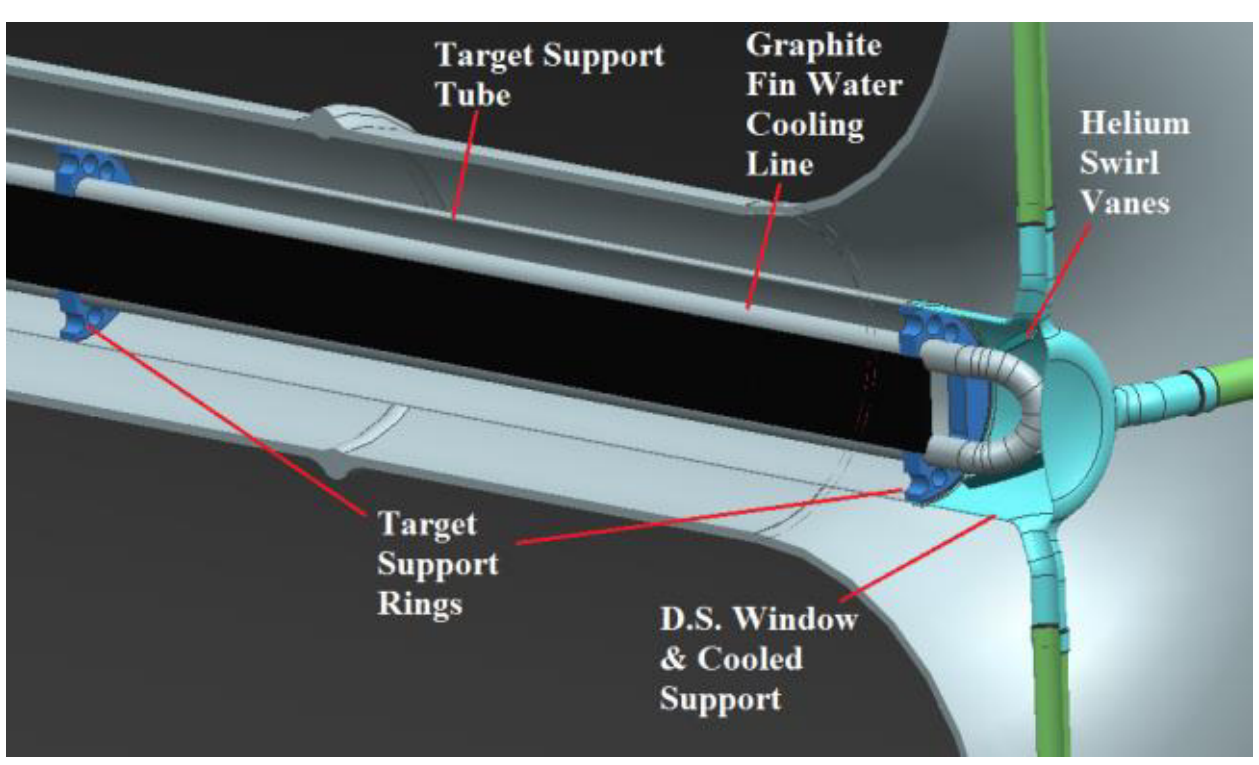

Figure 4: Revised target support tube and cooling system concept for integrated horn 1/target assembly.

Efforts to prototype the downstream window/support-tube transition piece must be undertaken to understand material properties and processing concerns for a production horn and target. Additional studies of fin temperature for target survivability, and radiative heat transfer from the target tube to the inner conductor are forthcoming.

### *Replaceable Beam Window*

Replaceable window design challenges include maximizing window lifetime when exposed to unprecedented secondary beam flux and developing the required remote handling system (see Fig. 5). Beryllium alloy will be used in the centre portion of the window and aluminium used in the outer portion. The use of beryllium reduces energy absorption, and thereby stress, to sustainable levels. Prototyping the beryllium to aluminium joining technology will be necessary. Seals used in this radioactive environment need to be radiation hard inorganic material. Clamping force requirements used for commercially available seals are very high, which constrains the window seal system design.

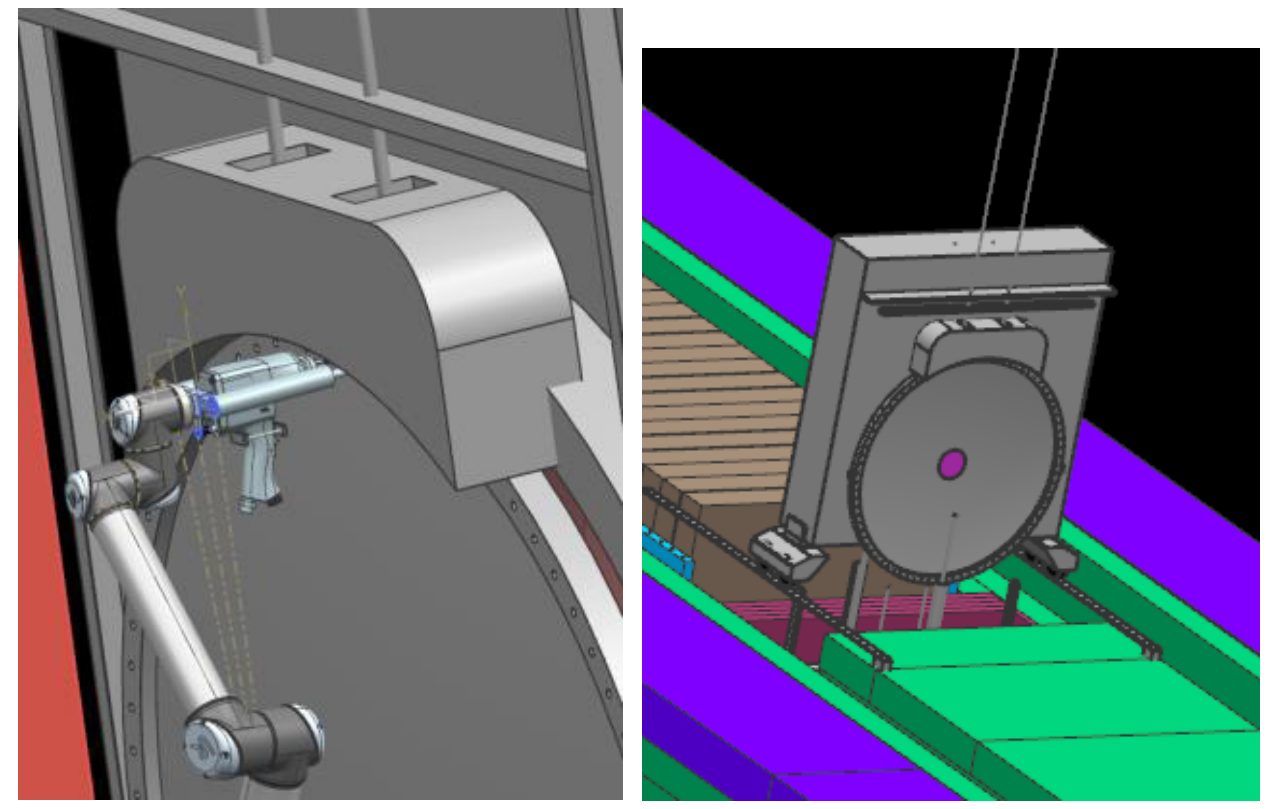

Figure 5: Replaceable beam window concept. The left figure shows a tooling concept for changing out the decay pipe window including a combination of a commercial six axis robotic arm and an impact wrench. The right figure shows a replacement decay pipe window assembly with a radiation shielding cask and helium air lock needed to efficiently exchange the decay pipe window.

## CONCLUSION

The status and challenges of the LBNF beamline Target Station design have been described above. Some alternative options for the target and horn designs that could enhance the physics capabilities of DUNE are being explored. Alternatives for the target chase cooling gas medium to minimize air emissions and any corrosion related issues are also being evaluated. Evaluation of these alternatives will be completed before baselining.

## ACKNOWLEDGMENT

We would like to thank the broader LBNF beamline team for their numerous contributions towards developing and advancing the beamline Target Station design. We also thank the DUNE collaboration for contributions to the beamline optimization and detailed studies of the neutrino fluxes.

## REFERENCES

[1] LBNF/DUNE Conceptual Design Report, Volume 3, Annex 3A, “Beamline at the Near Site,” pp. 1-231 (2015); `http://lbne2-docdb.fnal.gov/cgi-bin/ShowDocument?docid=10686`, June 25, 2015.

[2] P. Adamson *et al.*, “The NuMI Neutrino Beam,” *Nucl. Instrum. Meth.*, vol. A806, pp. 279-306, 2016.